# ONLINE UNSUPERVISED LEARNING FOR DOMAIN SHIFT IN COVID-19 CT SCAN DATASETS

*Nicolas Ewen\*, Naimul Khan\**

\* Electrical, Computer, and Biomedical Engineering, Ryerson University, Toronto, ON

## ABSTRACT

Neural networks often require large amounts of expert annotated data to train. When changes are made in the process of medical imaging, trained networks may not perform as well, and obtaining large amounts of expert annotations for each change in the imaging process can be time consuming and expensive. Online unsupervised learning is a method that has been proposed to deal with situations where there is a domain shift in incoming data, and a lack of annotations. The aim of this study is to see whether online unsupervised learning can help COVID-19 CT scan classification models adjust to slight domain shifts, when there are no annotations available for the new data. A total of six experiments are performed using three test datasets with differing amounts of domain shift. These experiments compare the performance of the online unsupervised learning strategy to a baseline, as well as comparing how the strategy performs on different domain shifts. Code for online unsupervised learning can be found at this link: https://github.com/Mewtwo/online-unsupervised-learning

## 1. INTRODUCTION

In this study we aim to determine whether unsupervised online learning can increase classification performance of convolutional neural networks (CNNs) on COVID-19 CT scan datasets [1]. We will explore the scenario where the target datasets have no annotations, and have slight domain shifts from the available training data. A strategy for unsupervised online learning for COVID-19 CT scans is proposed, and its performance is evaluated on three different test sets.

CNN models have been effective for classification on medical imaging datasets [2][3][4][5][6]. CNN models can be hard to train on medical imaging datasets because they require large amounts of annotated data. Large amounts of annotated data may not be available for a variety of reasons, including cost and availability of expert annotators [5][6]. Data augmentation, transfer learning, and self supervision are methods that can be used to help increase CNN performance when there is a small amount of annotated data [5][6][7], however, unsupervised learning methods are needed when there are no annotations [8].

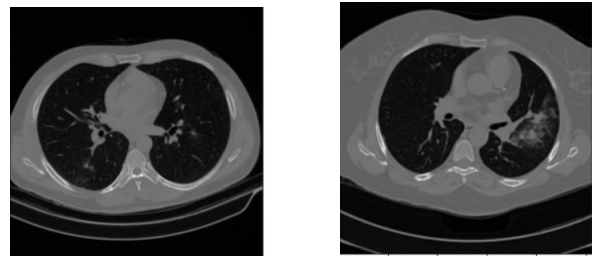

**Fig. 1**. Example of a CT scan with Covid-19 (left), and with CAP (right).

Performing transfer learning from a CNN pre-trained on ImageNet has improved model performance for COVID-19 CT scan datasets over training from scratch in a number of papers [9][4]. Some recent works further improved classification performance by using self supervision before transfer learning [8][10][11].

One method that is used to deal with few available annotations for training is transfer learning. Transfer learning is performed by taking a pre-trained network and then fine tuning it using the target dataset onto the target task [4]. This method allows simple and reusable features in the early layers of the network to benefit from training on an unrelated dataset. Since early layers are generally more reusable, the network will not have to train them as much, thus reducing the amount of annotated data needed to train the network [4][12].

Another method of dealing with a limited amount of available annotations for training is self supervised learning, which is a form of unsupervised learning. Self supervised learning uses unlabelled or automatically labelled data to pre-train the network to learn useful feature semantics in the target images [8]. Once the network has been pre-trained, it is then fine tuned onto the target dataset using transfer learning. This method allows the network to learn on limited labelled training data since less data will be needed after the network has already learned some useful feature semantics [8].

Online unsupervised learning is a combination of online machine learning and unsupervised learning. Online machine learning is a type of machine learning where the model continuously updates itself with new data as the new data arrives. Unsupervised learning allows models to learn from data without any expert annotations [8]. Online unsupervised learning allows the model to continuously improve as new data without annotations comes in [13].

Online unsupervised learning is a field of machine learning that can help predictive models adapt to new situations. New illnesses and screening methods, combined with a lack of expert annotators may cause domain shifts in incoming data, with no labels. Online unsupervised learning can help train models under such circumstances, where other techniques may have to wait for more data or annotations [13].

Our main contribution in this work is to highlight and demonstrate an online unsupervised learning strategy. While the idea of using unsupervised online learning to increase classification performance is not new [14], to the best of our knowledge it has not been done for medical imaging on a COVID-19 CT scan dataset.

We felt online learning was a good approach for COVID-19 CT scans because as more data is collected, the models can be updated. We decided to model what this might look like in practise by dividing each dataset into quarters and performing the online updates after each quarter. This would allow real-time results, while also continuously increasing classification performance.

The three test datasets are good to use for this experiment. The first test dataset comes from the same settings as the training and validation set. This set should be a benchmark for how much the model improves only due to extra training and data, since there is no domain shift. This means that with the first test set, we are performing semi-supervised learning, as opposed to unsupervised learning. The second test dataset is only COVID-19, and healthy patients, but with low dosage. This should be a slight domain shift. The third test dataset has COVID-19, CAP, and healthy patients. The patients also have a heart condition, and the dosage and slice thickness vary. This test set has the largest domain shift from the training and validation sets. Combined, the results from these three test datasets should show how well our proposed strategy of online unsupervised learning adapts to slight domain shifts.

This work is very important because if a method of unsupervised online learning can be used to increase classification performance under domain shift, then new models will not have to be trained from scratch each time the domain shifts slightly. It also means that a model can be updated as new data becomes available, without having to wait for an expert to annotate the new data.

We believe that our proposed strategy can be used in practise in the real world. Hospitals performing CT scans for COVID-19 can use their models to produce real time predictions, and the models update themselves as more data comes in. Since this strategy does not use annotations on the new data coming in, the model can be updated at a rate dependant on the rate of CT scans. As a demonstration of how this strategy can be used, we divide our three test datasets into four quarters. Each quarter will be treated as patients coming in sequential order. This means that the first quarter will be evaluated with our base models. The data from the first quarter will then be used to update the models. Then the second quarter will be evaluated using the updated models, and so on. By dividing our three test datasets into quarters, we are able to run six experiments to test how well this method of online unsupervised learning adapts to slight domain shifts in COVID-19 CT scan datasets.

## 2. METHODOLOGY

### 2.1. Dataset

The dataset we used for this paper was the dataset used in the SPGC COVID-19 competition [1]. This dataset is a dataset of chest CT scan images organized by patient. The patients can be in one of three classes: Healthy, COVID-19, or CAP. There are a total of 307 patients in the training and validation sets. There are 76 healthy patients, 171 COVID-19 patients, and 60 CAP patients. Patient-level labels are provided by three radiologists, who have greater than 90 percent agreement. Images were taken under different circumstances, including different medical centres, scanners, using different slice thicknesses, effective mA, and exposure time. 55 of the COVID-19 patients, and 25 of the CAP patients have slice-level labels provided by a single radiologist. There are about 5000 slices labelled positive for infection, and about 18,500 labelled negative for infection.

In addition to the training and validation sets provided, there are also three test sets. The first test set is from patients classified as healthy, COVID-19 positive, and CAP positive, and comes from the same distribution as the training and validation sets. The second test set from patients classified as healthy and COVID-19 positive only, and a lower dosage was used for the scans. The third test set comes from patients classified as healthy, COVID-19 positive, and CAP positive. The scans were administered under various settings. The healthy patients in this test set also had an unrelated disease.

### 2.2. Pre-processing

The SPGC COVID-19 dataset has pre-set training, validation, and testing sets. We used these sets and did not change them. We extracted the slice-level labels and images where slice-level annotations were provided. We kept the images and labels that were positive for COVID-19 and CAP, but we did not keep the negative ones. Instead we extracted slices from healthy patients with large lung area, in a similar manner to Rahimzadeh et al. [15]. We made an image selection algorithm to filter out images without lungs, or with small sections of lung, as well as images with lungs where much of the lung is not visible. This was done by setting an inner area of the image, and counting darker pixels in said area. A threshold was calculated for each patient on the fly using the average number of dark pixels, and images with less dark pixels in the area than the threshold were removed. These were used for normal slices. Images were resized to 224 x 224, and rescaled to between 0 and 1. The validation set was used to tune hyperparameters. For each patient, we also saved an array of the images with large lung area.

### 2.3. Slice-level models

We trained two different slice-level models. The first model was trained to classify slices as healthy or not healthy. This model was trained with the healthy slices we extracted from healthy patients, and the labelled slices provided for COVID-19 and CAP patients. The second model was trained to classify unhealthy slices as either COVID-19 or CAP. This model was trained with the labelled slices provided for COVID-19 and CAP patients.

For both models, we used the same network architecture and training strategy. The only difference was the data used. The network uses a DenseNet169 base, with a dense layer with 8 nodes followed by a softmax output. Batch normalization, regularization, and dropout were used as well. Our training strategy was a two-step process. For each model, we first performed targeted self supervision in a similar manner to Ewen and Khan [16]. We made horizontally flipped copies of our training images, and trained the network to determine whether an image was flipped or not. The second step in training was to then transfer onto our target dataset.

### 2.4. Patient-level models

At the patient level, we used the slices for each patient that we had previously extracted with larger lung area. The chosen slices were first sent through our slice-level model that classifies the slices as healthy or unhealthy. We took the average score of the patient's softmax scores to get two average scores: a healthy score, and an unhealthy score. If the healthy score was greater than five times the unhealthy score, the patient was classified as healthy. Otherwise, the patient was classified as unhealthy. This threshold of five times was chosen after testing on the validation set.

If the patient was classified as unhealthy, then the patients' slices were sent to the next slice-level model. The chosen slices were then classified as either COVID-19 or CAP. We again took the average softmax scores to get two average scores: a COVID-19 score, and a CAP score. If the COVID-19 score was greater than the CAP score, the patient was classified as having COVID-19. Otherwise, the patient was classified as having CAP.

### 2.5. Adjusting patient-level thresholds

On a number of runs, the slice level models produced a large difference in the recall of the classes. For example, due to many more images with COVID-19 than CAP, the slice level model could have a recall of about 0.99 for COVID-19 images, but only 0.72 for CAP images on the validation set. This suggests that the slice level model is more likely to classify CAP incorrectly than COVID-19. This could cause incorrect classification in borderline cases.

For example, if the slice level model classified 31 slices as COVID-19, and 30 as CAP, then the patient level model will more likely classify the patient as having COVID-19. However, since the patient only has one of either CAP or COVID-19, about 30 slices have been misclassified. Given the difference in the recalls of the slice level model, it is more likely that CAP images were classified incorrectly, and that it would be better to classify this patient as having CAP.

To deal with this problem, we came up with a method of adjusting the set threshold levels in the patient level models on the fly with a multiplier. To calculate these multipliers, we took the ratio of the two recalls. In our example this is CAP recall divided by COVID-19 recall. With a CAP recall of 0.72 and a COVID-19 recall of 0.99, this would be 0.72/0.99 = 0.725. We then multiply the number of COVID-19 slices by the multiplier. This would give us 31*0.725 = 22.5 effective slices. Since this number is lower than the 30 CAP slices, the patient is then more likely to be classified as having CAP. Similarly, a second threshold multiplier was also calculated for the healthy threshold.

If performances of the respective recalls are reversed, the threshold multiplier will be greater than 1. This means that the threshold adjustment will always be made in favour of the class with lower recall, and that as the difference in recall grows, the threshold multiplier, and therefore adjustment, gets larger.

### 2.6. Online Unsupervised learning

We tested this method using three different COVID-19 CT scan test sets. Our "baseline" contains two networks trained on the training and validation sets, first one providing healthy/unhealthy binary slice level classification, second one providing COVID/CAP slice classification on the unhealthy slices from the first model. For each test dataset, we retrained the slice level models using our online unsupervised learning scheme. This gave us 4 total models, the baseline model, and a model specifically fine tuned to its respective test dataset.

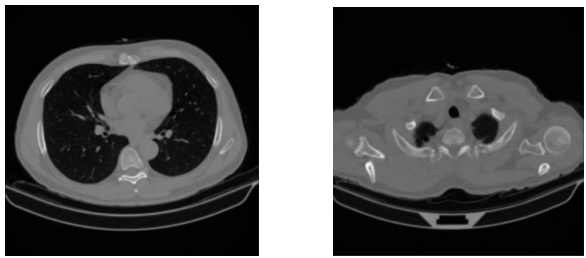

**Fig. 2**. Example of a healthy patient's slices with large lung area (left) and without (right).

Table 1. Experiments

|  | Proposed Experiments | |
|---|---|---|
|  | *Test Set* | *Model* |
| Exp. 1 | Test Set 1 | Baseline |
| Exp. 2 | Test Set 2 | Baseline |
| Exp. 3 | Test Set 3 | Baseline |
| Exp. 4 | Test Set 1 | Online Unsupervised |
| Exp. 5 | Test Set 2 | Online Unsupervised |
| Exp. 6 | Test Set 3 | Online Unsupervised |

To perform the online unsupervised updates to the models, we first ran the next quarter of test data through our models to obtain predictions. From these predictions we took the confident slices. Confident slices were those with a softmax score of at least 0.9 in agreement with the patient's classification. We then used these slices, along with our label that we assigned to it during predictions, and our original training data, to retrain the model. The aim was for this to allow the model to adjust to slight domain shifts, such as lower dosage.

We used a strategy that is adjusted from what was proposed by Cao and He [14]. After predictions were generated for a batch, a new slice-level model was trained for both healthy vs unhealthy classification as well as COVID-19 vs CAP classification. These two new models were initiated from the point that the self supervision step had finished for the base models. The confident images from the test batch, along with their predictions, were added to the original training and validation sets, and then trained in the same way as the base slice-level models. This means that after receiving every test batch, two new slice-level models were trained.

### 2.7. Experiments

We ran six experiments in total. For each test set, we ran an experiment using the baseline method to predict the class of the patients, and another experiment using the online unsupervised method to predict the patients' class. There are three test sets, so this resulted in a total of six experiments.

### 3. RESULTS

The results of the experiments can be seen in Table 2. On the first test set, the baseline method got 90 percent accuracy, while the online unsupervised method got 86.7 percent accuracy.This result was unexpected, since test set 1 comes from the same distribution as the training and validation set, and the baseline performed well. A possible explanation for this result is that the test set was too small. 30 patients may not be sufficient to demonstrate the proposed method. The initial guesses from the online method are the same as for the baseline, since the models have not yet updated. This means that model performance decreased even though most patients were correctly classified initially.

On the second test set, the baseline method got 66.7 percent accuracy, and the online unsupervised method got 76.7 percent accuracy. This dataset had a small domain shift from the training and validation sets, and seemed ideal for our method. The increase in performance is promising.

The third test set had the larger domain shift from the training and validation sets. The baseline method got 63.3 percent accuracy, while the online unsupervised method got 53.3 percent accuracy. This result is not entirely unexpected, as this test set had the largest domain shift. Another possible explanation for the poor performance on this test set is that the image selection algorithm may not be well suited to images from patients with heart conditions, as it may throw out useful images. A number of patients in this test set were left with significantly fewer images after the selection process compared to patients from the other test sets.

### 4. CONCLUSIONS

In this paper we demonstrated an online unsupervised learning method to boost performance of a COVID-19 classification model when tested with data with a domain shift from the training and validation sets. The aim of this method is to allow a model to adapt to a small domain shift in the data, without the need for expert labels.

Given the results of the experiments, we conclude that an online unsupervised learning method may be able to boost classification performance of COVID-19 diagnosis models under slight domain shift. However, further fine tuning is needed to see how much it can boost performance. Further explorations using different image selection algorithms may help boost performance on test set three. Testing on larger datasets may help clarify some of the current issues.

Table 2. Results of Experiments

|  | *Test set* | *Model* | *Accuracy*[a] |
|---|---|---|---|
| Exp.1 | Set 1 | Baseline | 0.9 +- 0.107 |
| Exp.2 | Set 2 | Baseline | 0.667 +- 0.169 |
| Exp.3 | Set 3 | Baseline | 0.633 +- 0.172 |
| Exp.4 | Set 1 | Online Unsupervised | 0.867 +- 0.122 |
| Exp.5 | Set 2 | Online Unsupervised | 0.767 +- 0.151 |
| Exp.6 | Set 3 | Online Unsupervised | 0.533 +- 0.179 |

[a]Accuracy with a 95 percent confidence interval.

## 6. ACKNOWLEDGEMENTS

We acknowledge NSERC's funding through an Alliance grant to conduct this study.